\documentclass[final,leqno]{article}
\usepackage[section]{placeins}
\usepackage[utf8]{inputenc}
\usepackage[english]{babel}
\usepackage{amsmath,amsfonts,amssymb,amsthm}
\usepackage{paralist}
\usepackage{graphics} 
\usepackage{epsfig} 
\usepackage{graphicx}
\usepackage{subfig}
\usepackage{caption}
\usepackage{epstopdf}
\usepackage[colorlinks=true]{hyperref}
\usepackage{comment}
\usepackage{url}
\makeatletter
\g@addto@macro{\UrlBreaks}{\UrlOrds}
\makeatother
\usepackage{bm}
\usepackage{fancyhdr}
\usepackage{lscape}
\usepackage{cleveref}

\hypersetup{urlcolor=blue, citecolor=red}

\newtheorem*{proof*}{Proof}

\begin{document}

\title{The role of mobility and sanitary measures on Covid-19 in Costa Rica, March through July 2020}

\author{Luis A. Barboza, Paola V\'asquez, Gustavo Mery, Fabio Sanchez,\\ Yury E. Garc\'ia, Juan G. Calvo, Tania Rivas, and Daniel Salas}  

\maketitle

\begin{abstract}
The aim of this paper is to infer the effects that changes on human mobility had on the transmission dynamics during the first four months of the SARS-CoV-2 outbreak in Costa Rica, before community transmission was established in the country. By using parametric and non parametric detection change-point techniques we were able to identify two different periods where at least the trend and variability of new daily cases significantly changed. In order to combine this information with population movement, we use data from Google Mobility Trends that allow us to estimate the lag between the rate of new daily cases and each of the categories established by Google. The information is then used to establish an association between changes in population mobility and the sanitary measures taken during the study period. 
\end{abstract}

\section*{Introduction}

The severe acute respiratory syndrome coronavirus 2 (SARS-CoV-2) is a highly transmissible and pathogenic novel human coronavirus initially detected in Wuhan City, Hubei Province of China in late December 2019 \cite{zhu2020novel,hu2020characteristics}. This pathogen is the causative agent of a wide spectrum of clinical manifestations, going from asymptomatic cases, patients with mild symptoms and severe disease with life-threatening complications \cite{wiersinga2020pathophysiology,docherty2020features}. After the initial cases were reported in China, and accelerated by international and domestic travel \cite{lau2020association,zhao2020quantifying}, the SARS-CoV-2 virus spread rapidly within and across countries and continents leading to a global health crisis and an unprecedented worldwide response in an attempt to reduce the importation of cases, to interrupt chains of transmission, to protect vulnerable populations and to mitigate the burden on health care systems, economies and societies \cite{srpcovid66:online,hartley2020public}. 

In this ongoing effort, countries around the globe have implemented a range of public health and social measures adapted to the local social, political, economical and evolving epidemiological context. Lockdown-type interventions and different forms of business closures have been widely used, intended to limit population mobility and to avoid close contact with individuals outside people's family unit \cite{cowling2020public, wilder2020isolation}. However, as the pandemic has unfolded, the role of specific sanitary measures in controlling the spread of Covid-19 has been largely disputed \cite{haug2020ranking,hsiang2020effect,flaxman2020estimating}. The concurrence of countless epidemiological and societal variables makes determining the effect of every individual measure a nearly impossible task.

One way to partially determine whether sanitary measures or packages of measures had a significant impact in disease transmission is through establishing the association between changes in population mobility and variations in the speed of disease transmission, and later identifying which sanitary measures were associated with those changes in mobility. Costa Rica represents an atypical setting to test this assumptions. In this particular Central American country, an initial effective control of the pandemic allowed avoiding community transmission for the first four months of the pandemic -between March and July 2020- when the virus was rapidly spreading in neighbouring countries and almost across the whole Americas \cite{Coronavi72:online}. Moreover, what makes Costa Rica a unique setting to conduct such analyses is the fact that the country never imposed a total or even partial lockdown. Without a lockdown, vehicular circulation restrictions were combined with several common measures, such as restrictions to commercial activity or school closures \cite{SitioWeb91:online}. This situation allows us to test these hypotheses before community transmission was established and detection of new cases became more challenging. 

We then examine the impact that differences in population's mobility dynamics had on the number of the new daily-confirmed COVID-19 cases in Costa Rica, since its initial detection within the country on March 6, 2020 up to the confirmation by health authorities of community transmission on July 2, 2020. 

A key element for the analysis is to be able to identify periods where the general behavior of new daily cases varied significantly. In this effort, the detection of single and multiple {\it change-points} in the time series of new cases has been a way to infer the impact of interventions in different countries. For example, Dehning et al. combined Bayesian inference with compartmental models to identify plausible {\it change-points} on the Covid-19 spreading rate in Germany \cite{dehning2020inferring} and Jiang et al. used a novel combination of algorithms to detect multiple {\it change-points} in the series of confirmed cases and deaths of Covid-19 over more than 30 countries \cite{jiang2020time}. Recently, Coughlin et al. showed an interesting application of change-point detection in Covid-19 using information of new cases in 20 individual countries (excluding Costa Rica) and the European Union in an aggregated way \cite{COUGHLIN2020100064}. They were able to identify change-points in trend and variability using a Bayesian {\it change-point} model together with a B-spline procedure. For mobility data, as previous studies around the world \cite{yilmazkuday2020stay, nouvellet2021reduction, chan2020risk}, we utilize data from Google's Mobility Reports to capture changes at least in tendency and variation of population movement to different locations. The results are then analysed with the public health measures adopted during the study period. The purpose of the study is therefore two-fold. First, we attempt to determine the validity of the association between population mobility within the national territory and the speed of disease transmission. Second, we identify which measures were adopted when those changes in mobility occurred, in order to explore the effectiveness of such sanitary measures in the control of the disease.

\section*{Methods and Materials}

 
\noindent{\bf Data.} The analysis uses three different sources of information to study the association among epidemiological and mobility variables:
\begin{itemize}
	\item Epidemiological data. The number of daily confirmed COVID-19 cases in Costa Rica from March 6, 2020, to July 2, 2020, was obtained from the Ministry of Health \cite{SitioWeb91:online}. The data covers 4,023 laboratory-confirmed cases from the 82 municipalities across the country. On March 6, Costa Rica became the 89th country to confirm a Covid-19 patient within national territory, a 49 year old woman visiting from New York \cite{CASOCONF5:online}. After twelve days the country reported its first death due to the virus \cite{Primerfa29:online}. During the initial months of the pandemic, the growth of cases was relatively stable in Costa Rica, and the majority of patients had a known chain of transmission. Two months into the pandemic, the country was even highlighted as having the lowest Covid-19 case fatality rate in the region \cite{5reasons54:online}. However, by the second week of June, 2020, the daily new cases began a progressive growth. On July 2, 2020 with 4,024 total confirmed cases, a 14-day case notification rate of 40.3 cases per 100,000 population and 17 deaths, the Ministry of Health declared community transmission in the Greater Metropolitan Area after being unable to establish the epidemiological nexus to $65\%$ of all new positive cases detected over the last five days \cite{Minister26:online}. This region comprises about $50\%$ of the national population. 
	
	\item Google's Community Mobility Reports \cite{google_mob}. This data contains relative changes of mobility according to Google's applications with respect to the data observed on a certain baseline. The baseline day is the median value using a five-week period on January and February, 2020. The relative changes are computed using the following categories: Retail and Recreation (restaurants, shopping centers, theme parks, etc), Grocery and Pharmacy (grocery markets, drugstores, pharmacies, etc), Parks (national parks, beaches, public gardens, etc), Transit stations (public transit hubs), Workplaces, and Residential.
	
    \item Sanitary measures. Since the first reported case of Covid-19 in Costa Rica, an inter-institutional and comprehensive approach led by the Ministry of Health, the Social Security System and the National Commission of Emergencies has guided the country’s response to mitigate the impact of the Covid-19 pandemic. One of the first measures aimed to promote and facilitate physical distancing taken by Costa Rica took place on March 10, 2020 with the cancellation of all massive events and the instruction of working from home for all public institutions \cite{Gobierno87:online}. On the days to come, the country restricted the capacity of public meeting spaces to $50\%$, closed schools and universities, all air and land borders, bars, beaches, churches, gyms, theaters and cinemas \cite{SitioWeb91:online}. By March 23, 2020, with a total of 158 confirmed cases spread across 30 municipalities (from a total of 82), public health authorities imposed the first restriction to vehicle circulation, beginning with a nighttime restriction to circulate from 10pm to 5am, with the exception of emergency vehicles, taxi drivers and those with work justification that provided a signed letter by the employer \cite{Gobierno53:online}. The opening, closure, and allowed capacity for both, commercial activities and social gathering places, as well as the vehicle restrictions, have been adjusted throughout the pandemic based on the transmission dynamics, international guidelines and comprehensive analyses of each municipality's transmission risk \cite{SitioWeb91:online}. 

\end{itemize}  


\noindent{\bf Methods.} The estimation of time points where the series of new infected cases changed in terms of trend, variability or more distributional properties, can give an insight of the approximate days where a certain intervention measure, or a combination of several measures, impact the underlying behavior of the transmission dynamics. In order to estimate where those changes probably occurred, we use non-parametric and parametric {\it change-point} detection techniques, in particular the Change Point model for sequential multiple change detection \cite{Ross2014,ross2011nonparametric} and the sequential detection of phase changes through outlier identification \cite{Chen1993}.

A Bayesian structural time series model fitted with a Markov chain Monte Carlo (MCMC) sampling algorithm is then combined with data from Google's Community Mobility Reports to infer the cumulative difference between the observed and expected number of the projected cases after the change-points were detected. 
We use several hypothesis tests to apply the non-parametric procedure: Mood \cite{mood1954asymptotic}, Lepage
\cite{lepage1971combination}, and the well-known Kolmogorov-Smirnov, Mann-Whitney and Cramer-von-Mises tests.

After we determine a set of {\it change-points}, we attempt to combine the above information to infer the causal impact of the mobility preceding each {\it change-point}. Here we use a certain type of Hierarchical Bayesian model called Bayesian structural time-series model \cite{brodersen2015inferring} (BST) with a set of $K$ lagged covariates with significant association with respect to the dependent variable. The covariate lags can be determined by means of a cross-correlation analysis. Two advantages of this model are that: (1) it is able to keep the temporal association in the modeling process, without using the restrictive assumption of independence in observations and (2) the Bayesian approach combined with a state-space structure assures a natural way to propagate the uncertainty along the model hierarchy. In our case, the BST model can be written as:
\begin{align}\label{BST}
y_t&=\mu_t+x_t^T\beta+\sigma_y\epsilon_t \nonumber\\
\mu_t&=\mu_{t-1}+\sigma_{\mu}\eta_t
\end{align}

\noindent where $y_t$ are the new infections due to Covid-19 at day $t$, ${\scriptstyle x_t^T=(M^{(1)}_{t-l_1},M^{(2)}_{t-l_2},\ldots,M^{(K)}_{t-l_K})}$ are lagged mobility covariates with their respective lags $l_k$ ($k=1,\ldots,K$),\\ $\beta=(\beta_1,\ldots,\beta_K)^T$, $\sigma_y$ and $\sigma_{\mu}$ are parameters, $\mu_t$ is a latent variable indicating the trend behavior of $y_t$, and $\epsilon_t$ and $\eta_t$ are normally-distributed white noises.

We use an MCMC algorithm to fit the model according to \eqref{BST}, using the available data before each change-point occurred and assuming for the subsequent {\it change-point} that the data starts immediately after the previous {\it change-point} happened. The parameter set $\theta=(\beta,\sigma_y,\sigma_{\mu})$ is assumed to follow a spike-and-slab prior according to Equation 2.8 in \cite{brodersen2015inferring}. The MCMC is performed with the R package \verb|CausalImpact| \cite{brodersen2015inferring} where posterior samples of the observed series and model parameters $\theta$ are computed. For each model the time period can be divided in two different sets: a \textit{pre-intervention} period containing the time before the {\it change-point} occurs and where the MCMC process is performed and a \textit{post-intervention} period where the MCMC process gives estimates of the observed series that we then compare with real values in order to approximate the causal effect of interventions.

To explore associations between the variations detected in population's mobility and the sanitary measures adopted by the health authorities, we identify all set of measures imposed from March through June, 2020 just before the appearance of the Covid-19 community transmission in Costa Rica. For simplicity, we assume that each set of measures is represented by an indicator at a unique date, chosen as the midpoint of the period where the set of measures is valid. We then apply a MANOVA model among the four mobility series in Table \ref{Table_lags} and two temporal components: (1) a weekly seasonal effect and (2) a conditional effect of the set of measures given that the time of the previous set is taken as the start of the fitting set of observations. 

Finally, we apply a Pillai test to infer the effect of each individual set of measures given its preceding set, to quantify its conditional impact on the mobility series. In this way we were able to associate the set of measures with the mobility information, to deduce their qualitative impact on the number of cases of Covid-19.

\section*{Results}

\begin{figure}[htb!]
	\centering
	\includegraphics[scale=0.5]{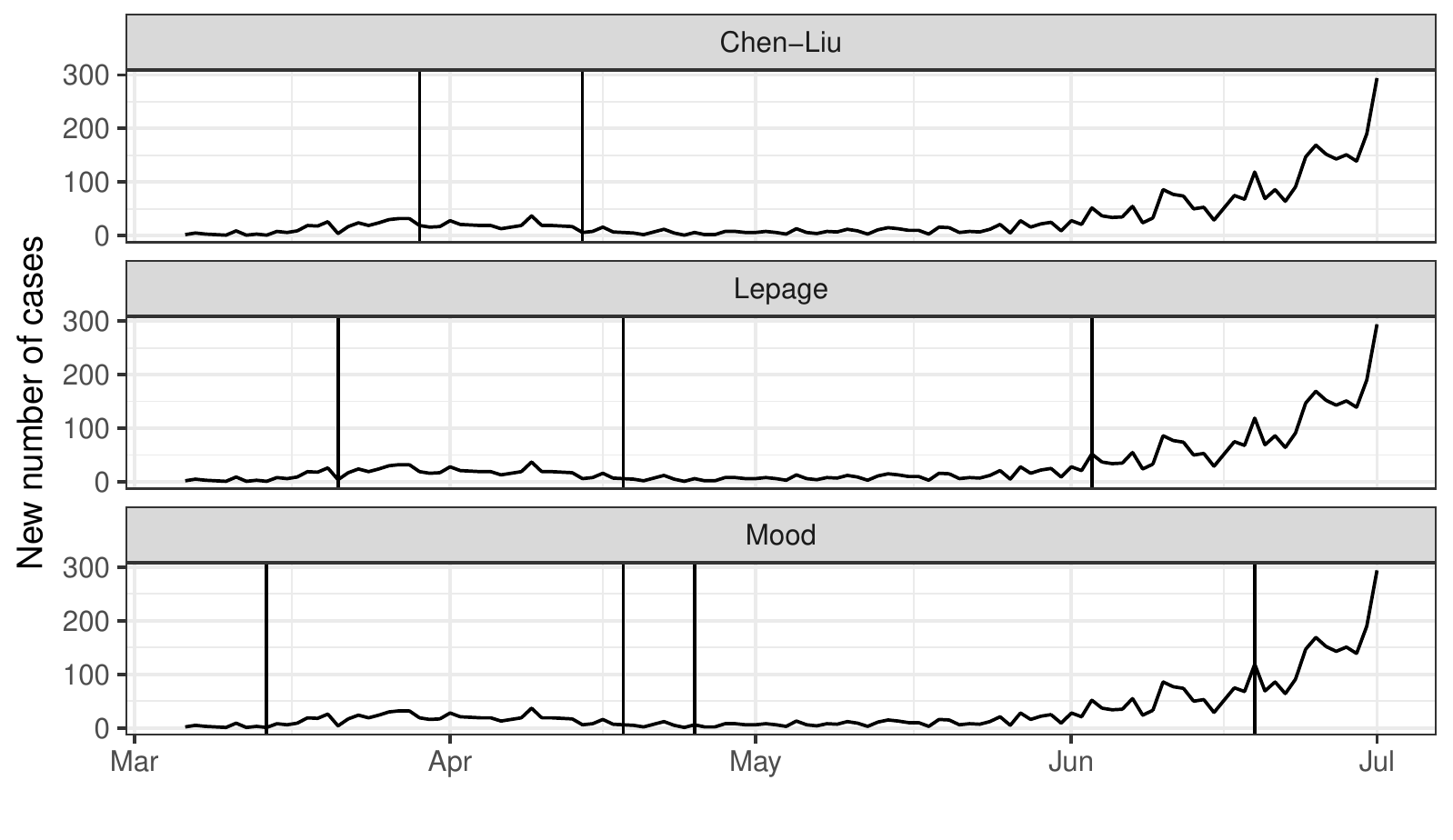}
	\caption{Estimated {\it change-points} of new Covid-19 infected cases for outlier identification (Chen-Liu) and sequential multiple {\it change-point} detection (Lepage, Mood) methods.}
	\label{fig:cppoints}
\end{figure}

Using the {\it change-point} model for sequential multiple change detection, we detected significant points with the non-parametric Mood \cite{mood1954asymptotic} and Lepage \cite{lepage1971combination} tests using the series of log-differences of new infected cases as a way to quantify the rate of change of the overall series. Figure \ref{fig:cppoints} shows the series of new infected Covid-19 cases with the estimated {\it change-points} with the available methodologies.  
Note that all the procedures were able to identify at least three different periods where the distributional properties of the new infected cases significantly changed. Those periods are not the same, but at least we have larger consensus among methods that the series of new infected cases experiences {\it change-points} in dates between April 14, 2020 and April 18, 2020 and between June 3, 2020 and June 19, 2020. For ease of analysis, we chose April 18, 2020 and June 19, 2020 as the estimated dates where the {\it change-points} occurred.

The changes experienced during mid-April and mid-June, 2020 are likely due to the impact of mobility restrictions or allowances on the days before the effect was quantified. A cross-correlation analysis among the rate of change of new cases and Google's Mobility Trends allowed us to determine which lags are more significant under a Pearson hypothesis test, as shown in Table \ref{Table_lags}. Based on the results from the cross-correlation analysis, the model used the number of daily confirmed cases starting on March 14, 2020, seven days after the first case was detected in the country.  
\begin{table}
	\centering
	\begin{tabular}{|c|c|c|}
	\hline
	\textbf{Categories}	& \textbf{Notation in \eqref{BST}} &\textbf{Lag (days)}\\
	\hline
	Retail and Recreation	& $M^{(1)}_t$ & 7 \\
	\hline 
	Parks	& $M^{(2)}_t$ & 7 \\ 
	\hline
	Transit stations & $M^{(3)}_t$	& 7 \\ 
	\hline
	Residential	& $M^{(4)}_t$ & 8 \\ 
	\hline
	\end{tabular} 
\caption{Significant cross-correlation lags under a Pearson test. The remaining categories show no-significant lags.}
\label{Table_lags}
\end{table}  

We then fit two BST models. The first one assumes that April 18, 2020 is a {\it change point} and the second one does the same with June 19, 2020 using the MCMC procedure with 1000 samples. We used as covariates the set of Google's mobility series shown in Table \ref{Table_lags} along with their lags. 

\begin{figure}[!ht]
\centering
\subfloat{\includegraphics[width=0.5\textwidth]{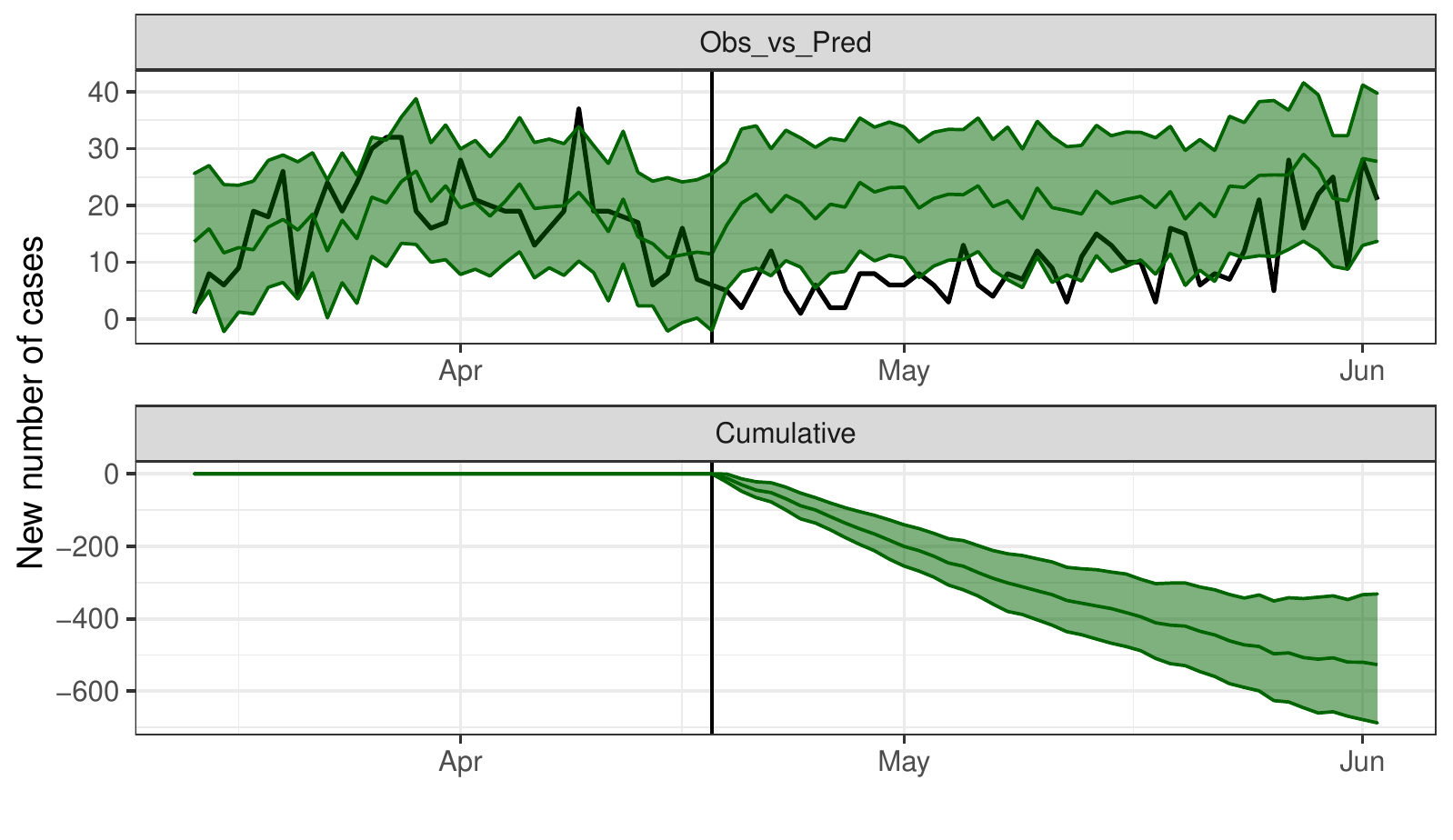}}\hfill
\subfloat{\includegraphics[width=0.5\textwidth]{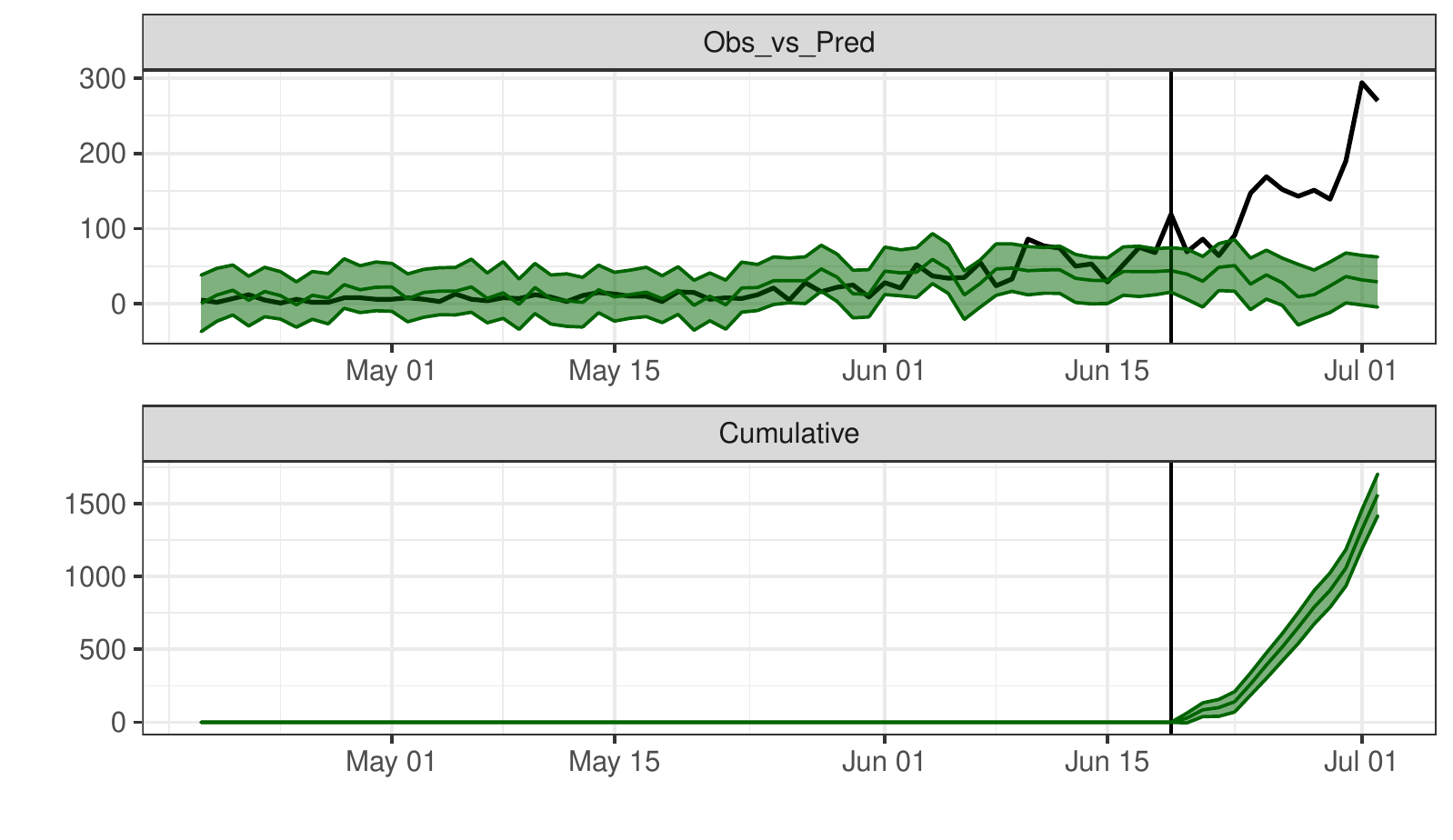}}\hfill
\caption{Upper panels: Bayesian estimation of the observed number of Covid-19 cases during two study periods: March 14-June 2, 2020 (left) and April 18-July 2, 2020 (right). Lower panels: estimated cumulative effect of the measures taken before April 18, 2020 (left) and June 19, 2020 (right).}
\label{fig:April18th}
\end{figure} 

The upper panels of Figure \ref{fig:April18th} contain the Bayesian estimation of the observed process for both periods together with their respective $95\%$ predictive regions. The lower panels contain the posterior estimation of the cumulative effect of the intervention along the post-intervention period.
 
\begin{figure}[htb!]
	\centering
	\includegraphics[scale=0.7]{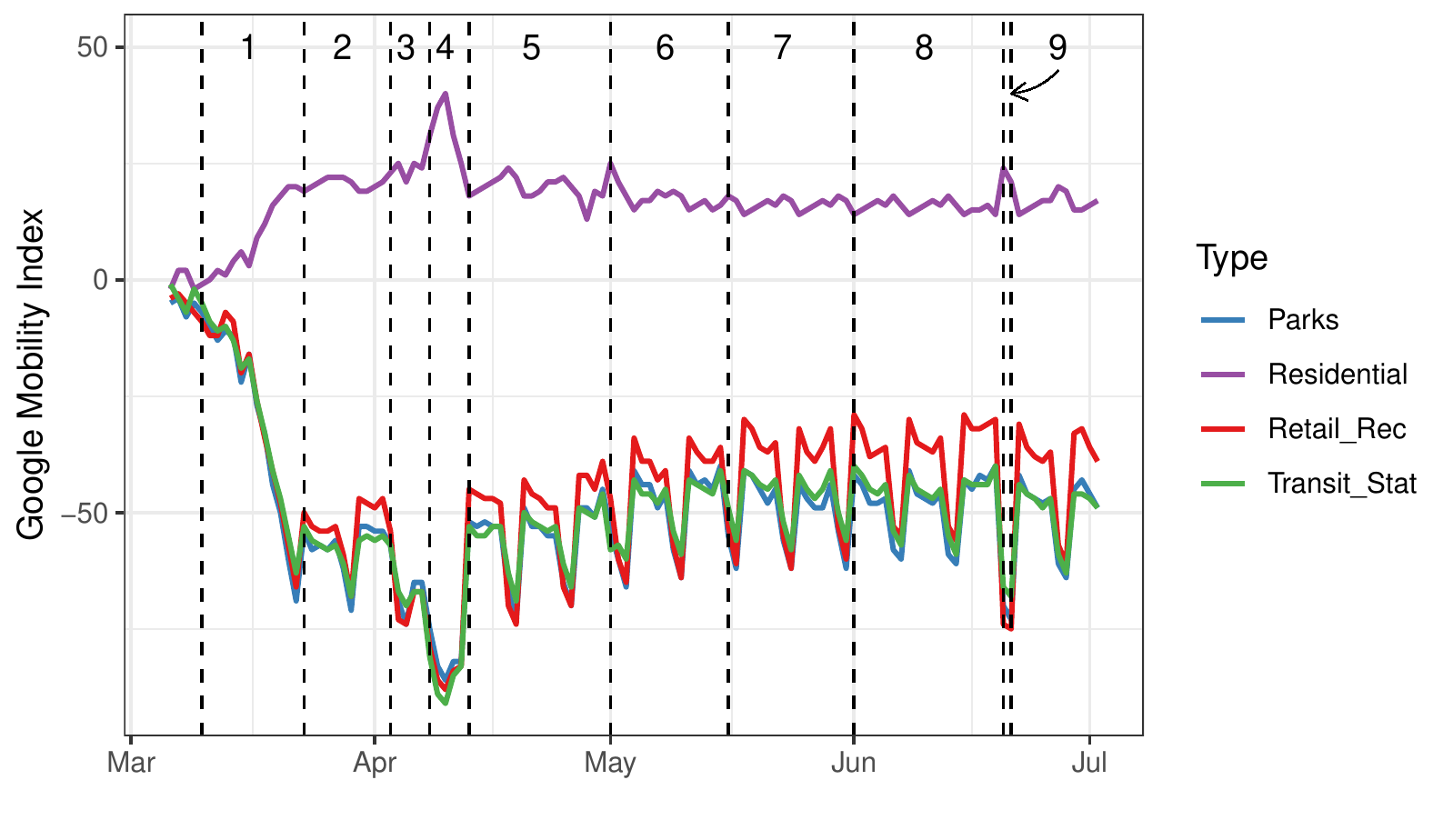}
	\caption{Google mobility indices (solid lines) along with the sets of sanitary and mobility measures (dashed lines). The sets of measures are defined in Table \ref{tab:measures}.}
	\label{fig:google_vs_measures}
\end{figure}  

The upper-left panel in Figure \ref{fig:April18th} shows the Bayesian estimation of the causal effect using March 14, 2020 through April 18, 2020 data as the pre-intervention period and April 19, 2020 through June 2, 2020 as post-intervention period. We remark that between 342 and 697 cases represent a $95\%$-confidence interval of the cumulative effect at June 2, 2020 of the set of interventions before April 18, 2020, with a mean value of 526 cases. This mean value approximately represents a $47\%$ of the observed cumulative cases at June 2, 2020. The daily difference among the expected (under the pre-intervention conditions) and the observed number of cases is between 6 and 15 cases with the same confidence level, with a mean difference of 11 cases. In relative terms, there is a mean reduction of $54\%$ in the number of daily cases with a confidence interval of $[35\%, 71\%]$ approximately.   

The upper-right panel in Figure \ref{fig:April18th} shows the fitting with April 19, 2020 through June 19, 2020 as pre-intervention period and June 20, 2020 through July 2, 2020 as post-intervention period. In this case the effect is the opposite. There is a mean cumulative difference between observed and estimated cases of 1965 cases on July 2, 2020. This represents a $48.8\%$ of the observed cumulative cases at same date. The daily difference among the observed and the expected number of cases (under the pre-intervention conditions) is between 109 and 131 cases with the same confidence level, with a mean difference of 120 cases. In relative terms, there is a mean increase of $387\%$ in the number of daily cases with a confidence interval of $[351\%,423\%]$ approximately.

Table \ref{tab:measures} contains the sets of measures from March 10, 2020 to June 21, 2020, just before the community transmission started in Costa Rica. The conditional effect of each set of measures is shown in the fourth column, measured by the p-value of Pillai test. Note that all the sets cause a significant effect on the group of mobility covariates when we take in account the weekly effect of mobility patterns. However, the effect of mobility with respect to sanitary measures varies across time periods. The first two sets are expected to show a larger effect because of the abrupt change in mobility behavior during March 2020 (see Figure \ref{fig:google_vs_measures}). Afterwards, the conditional effect of additional measures is smaller, but set 4 shows a larger impact due to the general mobility restriction during Holy Week in Costa Rica. Starting from set 5, the subsequent mobility openings had a significant impact on Google's indices and the effect tends to decline by the end of June 2020.

\newpage
\begin{landscape}
\begin{table}[]
\centering
\small
\begin{tabular}{|l|l|l|l|}
\hline
\textbf{Set} &
  \textbf{Date} &
  \textbf{Measures} &
  \textbf{Pillai test} \\ \hline
1 &
  3/10 - 3/22 &
  \begin{tabular}[c]{@{}l@{}} Cancellation of massive events and work from home instructions\\ Closure of bars, casinos\\ Government declares National Emergency. Closure of all schools and borders \\ Closure of movies, theaters, gymnasiums, malls at $50\%$ \end{tabular} &
  $< 2.2\times 10^{-16}$ \\ \hline
2 &
  3/23 - 4/2 &
  \begin{tabular}[c]{@{}l@{}} Closure of beaches, churches, national parks. Start of vehicle restrictions from 10pm to 5am \\  Vehicle restriction on weekends from 8pm to 5am \\ Closure of all commercial activities at 8pm and during the weekend\end{tabular} &
  $8.94\times 10^{-14}$ \\ \hline
3 &
  4/3 - 4/7 &
  \begin{tabular}[c]{@{}l@{}}Daytime vehicle restriction from 5am to 5pm with plate distribution and nighttime vehicle\\ restriction from 5pm to 5am.\\ Restriction to long-distance public transportation.\\ Closure of all non-essential commercial activities.\end{tabular} &
  $2.38\times 10^{-3}$ \\ \hline
4 &
  4/8 - 4/12 &
  The circulation of vehicles and public transportation was suspended. &
  $2.71 \times 10^{-11}$ \\ \hline
5 &
  4/13 - 4/30 &
  \begin{tabular}[c]{@{}l@{}}Day-time vehicle restriction from 5am to 7pm with plate distribution.\\ Total nighttime vehicle restriction from 7pm to 5am\\ On weekends total vehicle restriction (with exceptions to access to essential services)\\ Commerce may function from 5am to 7pm, weekends delivery only.\end{tabular} &
  $9.65 \times 10^{-10}$ \\ \hline
6 &
  5/1 - 5/15 &
  \begin{tabular}[c]{@{}l@{}}The first phase of the gradual reopening of commercial activities.\\ Vehicle restriction from 5am to 7pm with plate distribution. Total nighttime restriction\\ from 7pm to 5am.\end{tabular} &
  $2.01 \times 10^{-7}$ \\ \hline
7 &
  5/16 - 5/31 &
  \begin{tabular}[c]{@{}l@{}}The second phase of the gradual reopening of commercial activities.\\ Vehicle restriction from 5am to 10pm with plate distribution, total nighttime restriction\\ from 10pm to 5am.\end{tabular} &
  $2.03 \times 10^{-5}$ \\ \hline
8 &
  6/1 - 6/19 &
  \begin{tabular}[c]{@{}l@{}}The third phase of the gradual reopening of commercial activities.\\Vehicle restriction from 5am to 10pm with plate distribution, total nighttime restriction\\ from 10pm to 5am.\\ Differentiated vehicle restriction in municipalities located near the northern border area\\ of Costa Rica.\end{tabular} &
  $8.84 \times 10^{-4}$ \\ \hline
9 &
  6/20 - 6/21 &
  \begin{tabular}[c]{@{}l@{}}Total vehicle restriction with circulation only to essential services \\ Closure of commercial activities\end{tabular} &
  $8.84 \times 10^{-4}$ \\ \hline
\end{tabular}
\caption{Sanitary and mobility measures from March through June, 2020. The p-value of the Pillai test is shown in the fourth column.}
\label{tab:measures}
\end{table}
\end{landscape}

\section*{Discussion}

As the transmission of the SARS-CoV-2 virus has progressed around the world, the primary focus of decision makers has been to implement comprehensive public health measures adapted to each country's unique capacities and context \cite{covidstr47:online}. With the beginning of a new phase in the Covid-19 pandemic (as vaccination programs become available, the emergence of new strains \cite{InterimI25:online,mahase2021covid}, and behavioral patterns and physical contacts constantly evolving), the necessity of balancing the social and economic impact of sanitary measures, and understanding their effects on the overall transmission in specific settings, are both of paramount importance. 

For Costa Rica, the current study has demonstrated an association between sanitary measures and variations in human mobility patterns, influencing the transmission dynamics of Covid-19 in the country. Through the use of {\it change-points} algorithms, we were able to divide the time series data corresponding to the number of confirmed cases from March 14, 2020 to July 2, 2020 in two periods where the tendency of cases significantly varied. After the first detected {\it change-point} on April 18, 2020, the models were able to estimate a mean reduction of $54\%$ in the number of daily cases from the projected number. Based on the estimated lagged period of seven days and the high correlation estimated by the Pillai test, this reduction in the tendency of cases can be traced back to the set of sanitary measures taken by the health authorities during Holy Week, from April 3, 2020 to April 12, 2020 \cite{Enelmarc86:online}. In an effort to reduce human mobility during a holiday period for most of the population, the Ministry of Health announced sanitary measures that were divided in two phases. The first one from April 3, 2020 (Friday) to April 7, 2020 (Tuesday), which involved the total restriction of vehicle circulation from 5pm to 5am, the regulation of vehicle circulation by license plate from 5am to 5pm, the restriction to long-distance public transportation for distances longer than 75 kilometers, and the closure of commercial activity that involves attention to face-to-face public. The second set of measures from April 8, 2020 (Wednesday) to April 12, 2020 (Sunday), has been the most restrictive one that Costa Rica has taken throughout this ongoing pandemic. It included the total restriction of vehicle circulation, suspension of public transportation, and limited access to commercial activities (only essential services). During March and April 2020, the epidemiological context of the country was led by well established clusters of cases and a slow increase in the number of daily cases. However, the country was also in a race to increase hospital capacity, mainly for intensive care units. This health crisis situation drove the Ministry of Health to implement restrictive measures in order to flatten the curve of new cases, to avoid the saturation of health care services and to give time to health officials to increase the human resources, hospital beds, ventilators and personal protection equipment necessary to care for the patients. 

The epidemiological context and results of the second detected {\it change-point} was very different from the first period. The analyses showed an increase of $387\%$ in the number of new daily cases with respect to the projection by the Bayesian analyses. Before June 20, 2020, the country started the gradual reopening of commercial activities, as the number of new cases were stabilizing and health officials had started to increase hospital capacity. These measures started on May 1, 2020 \cite{Gobierno63:online}, divided in periods of 14 days each and involved the gradual opening of businesses such as gymnasiums, movie theaters, beauty salons, hotels, as well as the opening of beaches with restrictive schedules, and national parks \cite{Apartird42:online}. On June 20, 2020, the country was set to start a new phase of reopening, which coincided with the celebration of Father's day in Costa Rica. However, due to the increase in the number of new cases, on June 19, 2020, the same date detected as {\it change-point}, the Ministry of Health announced the suspension of all reopening activities, and instead implemented the total restriction of vehicle circulation during that weekend (June 20-21, 2020) \cite{Terceraf73:online}. Besides the gradual reopening of commerce, the country was also starting to witness an increase in the number of new cases detected in municipalities located near the northern border of Costa Rica. The cases were mostly linked with agricultural activities, with clusters being reported in locations such as processing factories \cite{Anteincr33:online}. The incidence increase in that particular area of the country led to the implementation of differentiated vehicle restrictions for the region. On the first weeks of June, 2020, health authorities were also reporting an increase in the number of new cases related to social activities \cite{Establec29:online}. Overall, during this second period of the analysis, although mobility dynamics were positive correlated with the implementation of the set of public health measures, the results of the Pillai test showed a weaker conditional effect which can be expected due the complex interaction of the other social, economical and behavioural factors previously discussed. Therefore, limiting the capacity of the models to accurately associate the increase in daily new cases with the public health measures implemented during that period. 

We remark that one of the limitations of the study is that mobility data is aggregated at a national level. Therefore, it was impossible to analyse differences of mobility by region. Moreover, the statistical analyses relied on a series of assumptions. For example, the use of a single midpoint to represent each set of measures can change their real effect over the mobility series, or the effect not accounted by the mobility data on the detected {\it change-points} can have more temporal structure than the one assumed in Equation \eqref{BST}.

Despite these limitations, results from the study provide an initial effort to quantify the changes in transmission dynamics of Covid-19 during different periods in Costa Rica. 

\section*{Acknowledgements}
The authors would like to thank the Research Center in Pure and Applied Mathematics and the School of Mathematics at Universidad de Costa Rica for their support during the preparation of this manuscript. They also thank the Ministry of Health for providing data and valuable information for this study. Thanks to Dr. Mar\'ia Dolores P\'erez-Rosales, Representative of the Pan American Health Organization/World Health Organization in Costa Rica for her support and encouragement to pursue this work.

\FloatBarrier

\end{document}